\documentclass[journal]{IEEEtran}
\usepackage{subfigure}
\usepackage{graphicx}
\usepackage{bm}
\usepackage{amsmath}
\usepackage[table]{xcolor}
\usepackage{caption}
\usepackage{tabularx}
\usepackage{booktabs}
\usepackage{array}
\ifCLASSINFOpdf
\else
\fi

\usepackage{xcolor}

\begin{document}
\bstctlcite{BSTcontrol}

\title{Wave-Domain Semantic ISAC: Holographic Surfaces as Analog Semantic Encoders for Integrated TN-NTN}

\author{Shadi Danesh, ~\IEEEmembership{Member,~IEEE},  Pooria Salami,  
Hesaam Zaravashan,  
Abdulkadir Köse,~\IEEEmembership{Member,~IEEE}, Ayhan Akbas,~\IEEEmembership{Member,~IEEE}, Chuan Heng Foh,~\IEEEmembership{Senior~Member,~IEEE}, and Mohsen~Khalily,~\IEEEmembership{Senior~Member,~IEEE}
        
\thanks{ S. Danesh, P. Salami, H. Zaravashan, C. Foh and M. Khalily are with the Institute for Communication Systems (ICS), 5GIC \& 6GIC, University of Surrey, Guildford, Surrey, GU2 7XH, U.K. Email:\{s.danesh, p.salami, h.zaravashan, c.foh, m.khalily\}@surrey.ac.uk.}

\thanks{ Abdulkadir Köse is with the Department of Computer Engineering, Abdullah Gül University, Kayseri, Türkiye. Email: abdulkadir.kose@agu.edu.tr}

\thanks{ Ayhan Akbas is with the Computer Engineering Dept.
Ankara Bilim University, Ankara, Türkiye. Email: ayhan.akbas@ankarabilim.edu.tr}

}

\maketitle

\begin{abstract}
Semantic communication, near-field operation, and integrated sensing and communication (ISAC) are often considered separately in terrestrial and non-terrestrial networks (TN--NTN). We show how a large electromagnetic aperture can connect them.  \textcolor{black}{Nonlocal metasurfaces and holographic apertures act on the spatial spectrum of an incident field and can execute linear projections, which form an early stage of semantic feature extraction, directly in the wave domain before digitisation, thereby reducing the corresponding digital processing burden.} Near-field focusing also provides position-dependent sensing features, allowing the aperture to serve as an analog semantic-ISAC front end on resource-constrained aerial and satellite platforms. We describe the underlying physics, the co-design of constrained analog projections with lightweight digital decoders, and a case study in which a data-adaptive projection outperforms a fully digital baseline in the simulated energy-limited regime while transmitting four times fewer coefficients and using 11.4 times fewer decoder MACs. Finally, we identify relevant 3GPP, ITU-R, and ETSI activities and the remaining standardisation gaps.
\end{abstract}


\IEEEpeerreviewmaketitle

\section{Introduction}

Integrated terrestrial and non-terrestrial networks (TN-NTN) are a central part of the sixth-generation (6G) vision. The 3GPP NTN study established the foundation for incorporating satellites, high-altitude platform stations (HAPS), and uncrewed aerial vehicles (UAVs) into New Radio (NR)~\cite{rf1}, while the ITU-R IMT-2030 framework identifies ubiquitous coverage, integrated sensing and communication (ISAC), and artificial intelligence as key capabilities~\cite{ITU2160}. However, low-Earth-orbit (LEO) payloads, solar HAPS, and small UAVs face tight power, thermal, processing, and feeder-link constraints. Onboard computation competes directly with radio-link resources, while forwarding raw sensor data consumes scarce backhaul capacity.

Semantic communication can reduce this burden by conveying task-relevant information instead of a faithful source reconstruction~\cite{Shannon1949,gunduz2022beyond,strinati20216g}. Yet conventional semantic encoders are deep neural networks executed in digital baseband, where they consume the same limited computational and thermal resources they are intended to save. Thus, the platforms that benefit most from semantic communication may be least able to afford its conventional implementation.
This article addresses that paradox using large electromagnetic apertures. 

\begin{figure*}
    \centerline{\includegraphics[width=6 in]{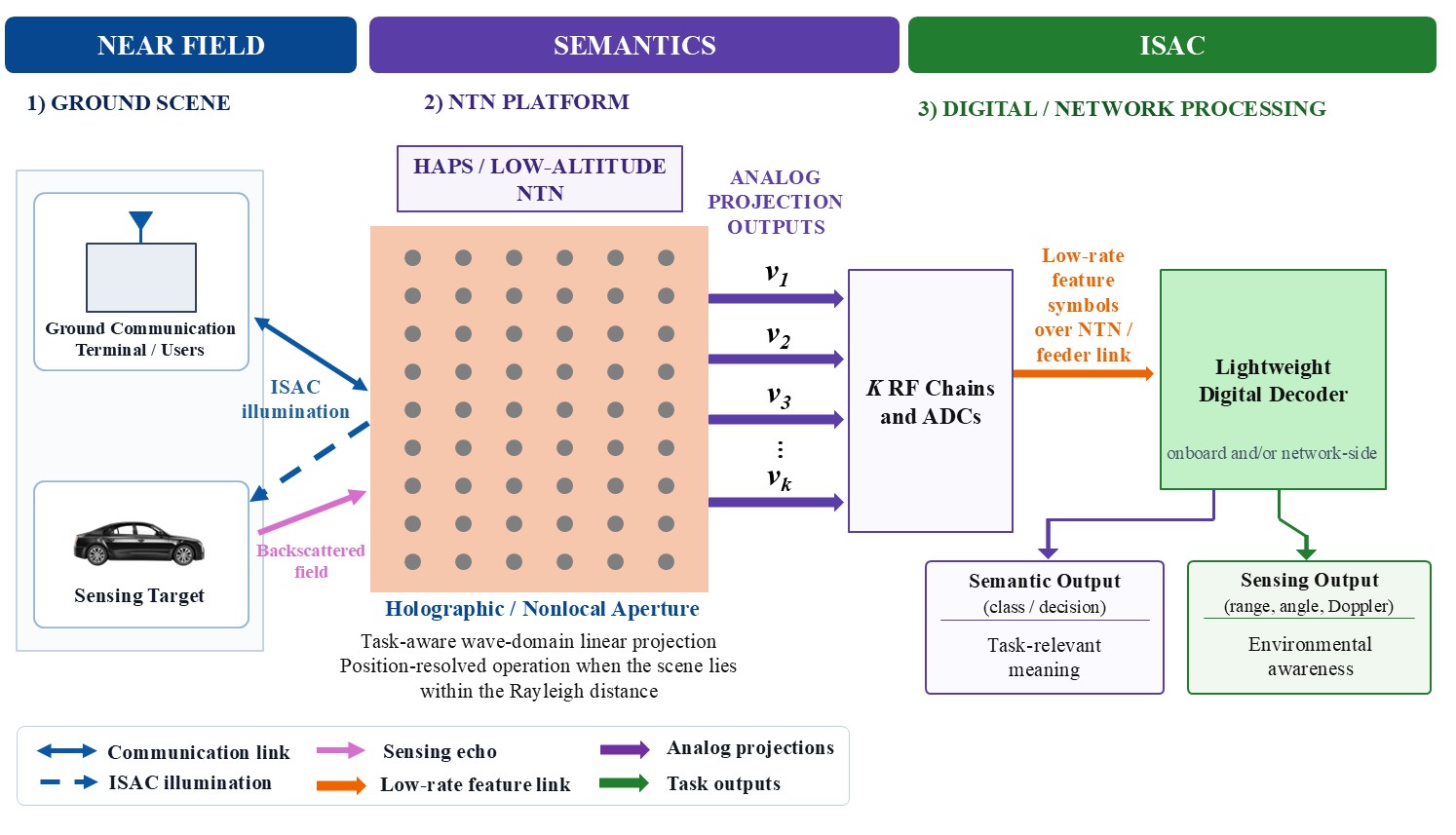}}
    \caption{\textcolor{black}{Conceptual wave-domain semantic ISAC architecture for integrated TN-NTN. Incident communication and sensing fields form the input to the task-aware holographic or nonlocal aperture. Its analog projection outputs are conditioned by RF chains, digitised by analog-to-digital converters (ADCs), and supplied to a lightweight decoder that produces task-level communication and sensing outputs.}} 
    \label{fig:architecture}
\end{figure*}

At millimetre-wave and sub-terahertz frequencies, selected TN--NTN links involving sufficiently large apertures or comparatively low-altitude platforms can operate in the radiative near field~\cite{liu2023near}. \textcolor{black}{In this regime, wavefront curvature encodes both range and angle in the spatial-field distribution across the aperture, providing position-dependent information that can be processed before digitisation.} Nonlocal metasurfaces respond selectively in spatial frequency, while holographic apertures can compute overlaps between incident fields and stored patterns~\cite{huang2020holographic,di2020smart}. They can therefore act as hardware-constrained linear projectors, executing the initial filtering or projection stage of a semantic encoder before sampling and digitisation. \textcolor{black}{This avoids the corresponding digital multiply-accumulate operations at the projection stage, although the remaining readout, conversion, calibration, control, and decoding stages still consume energy.} The concept extends metamaterial analog computing~\cite{silva2014performing,zangeneh2021analogue} from mathematical field operations to semantic feature extraction. 
\textcolor{black}{The same spatial-processing capability can also support sensing. In an ISAC system, communication and sensing share waveforms, hardware, and spectral resources~\cite{liu2022integrated}. The aperture can therefore process communication signals together with sensing-related information, including spatial features associated with range and angle and temporal features such as micro-Doppler, all of which provide task-relevant inputs to the downstream semantic processor. As illustrated in Fig.~\ref{fig:architecture}, the shared aperture transforms the available communication and sensing information into a compact set of task-relevant coefficients before digitisation and lightweight digital decoding. In this way, near-field processing provides the physical mechanism for manipulating spatial information, semantic processing determines which information is useful for the task, and ISAC provides the system-level framework for using a shared aperture for communication and sensing.}


This article makes three contributions. First, it explains how nonlocal and holographic apertures implement linear semantic projections, use near-field spatial structure for position-dependent feature extraction, and operate within physical constraints. Second, it presents the co-design of a constrained analog front end and a lightweight digital decoder, supported by a case study in which a data-adaptive projection outperforms a fully digital baseline in the simulated energy-limited regime while transmitting four times fewer coefficients and requiring 11.4 times fewer decoder multiply--accumulate (MAC) operations. Third, it maps the proposed architecture onto current 3GPP, ITU-R, and ETSI activities and identifies the remaining specification gaps.

\textcolor{black}{The end-to-end processing chain comprises four stages. First, the communication and sensing scene produces an electromagnetic field over the aperture. Second, the holographic or nonlocal surface transforms this field into a compact set of task-oriented outputs. Third, the radio-frequency (RF) chains and converters sample and digitise only these outputs. Finally, a lightweight digital processor maps the resulting features to communication, sensing, or task-level decisions. The aperture therefore performs the initial task-oriented preprocessing stage rather than replacing the digital processor.}


\section{Three Musketeers, One Aperture}

\textcolor{black}{Shannon and Weaver distinguished the technical problem of reproducing symbols from the semantic problem of conveying meaning~\cite{Shannon1949}. Semantic communication builds on this distinction by transmitting the information required by the receiving task rather than preserving every source detail~\cite{gunduz2022beyond,strinati20216g}. Its conventional implementation employs a learned encoder to produce a compact semantic representation and a decoder to recover the task output. However, the encoder is normally executed by a digital processor colocated with the transmitter, which may be unsuitable for power-, processing-, and thermally constrained non-terrestrial network (NTN) platforms.}

\textcolor{black}{Near-field operation occurs within the Rayleigh distance of an electrically large aperture, where wavefront curvature makes the channel response dependent on both range and angle~\cite{liu2023near}. The aperture can therefore focus on a spatial position rather than only in a direction. Beyond improving link efficiency, this position dependence has semantic value: projection bases constructed from focusing kernels can resolve scene geometry relevant to NTN tasks such as localisation, traffic monitoring, and platform coordination. Near-field operation also exposes richer spatial structure across the aperture for subsequent wave-domain processing.}

\textcolor{black}{Integrated sensing and communication (ISAC), the third ingredient, enables communication and sensing to share spectrum, hardware, and waveforms~\cite{liu2022integrated}. The 3GPP Release~19 study defines associated use cases, service requirements, and information-exposure considerations~\cite{3GPP_TR_22_837}. In a semantic architecture, communication coefficients and sensing products, such as range profiles or compressed spatial observations, can both serve as task-relevant feature vectors. A shared computational aperture can therefore project communication and sensing fields into a common compact representation for downstream processing.}

\section{The Physics of Wave-Domain Semantic Extraction}
\label{sec:Physics of Wave-Domain}

\textcolor{black}{Wave-domain semantic extraction uses the physical response of an electromagnetic aperture to transform an incident field into compact, task-relevant features before digitisation. In the proposed architecture, this processing is realised through nonlocal spatial-frequency filtering and holographic projection, while near-field wavefront curvature provides position dependence and physical constraints limit the achievable transformations.}

\subsection{Nonlocal Surfaces as Spatial-Frequency Filters}

\begin{figure*}
    \centerline{\includegraphics[width=7 in]{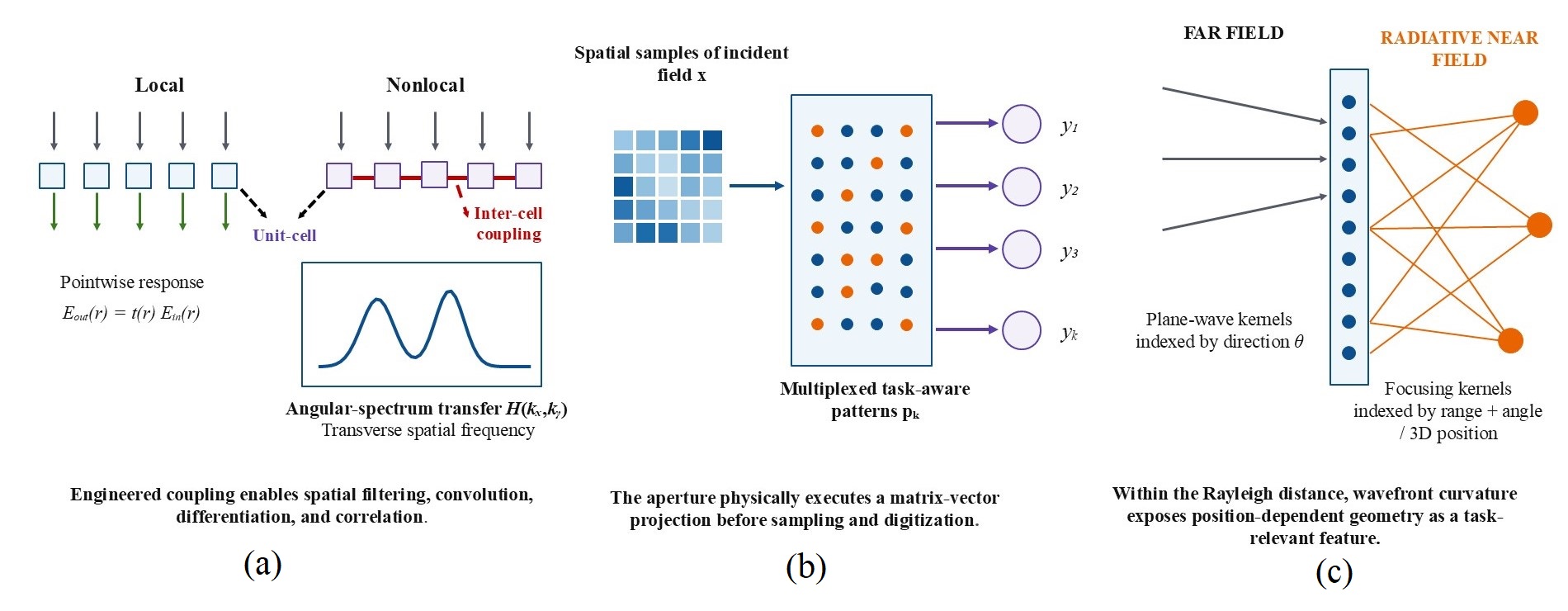}}
    \caption{\textcolor{black}{Conceptual progression from electromagnetic spatial processing to task-oriented features. (a) An incident field is processed locally through pointwise responses or nonlocally through an angular-spectrum transfer function. (b) Spatial samples of the incident field are projected onto multiplexed task-aware patterns to produce a reduced feature vector. (c) Near-field focusing replaces direction-only far-field kernels with kernels indexed jointly by range and angle. The content is explanatory and is not a full-wave simulation result.}}
\label{fig:wave_domain_physics}
\end{figure*}

\textcolor{black}{Conventional phased arrays and local metasurfaces primarily apply controllable element-wise or pointwise responses for functions such as beam steering and focusing. A nonlocal metasurface, by contrast, is a physical surface whose spatially dispersive response arises from engineered interactions among neighbouring elements, guided-mode resonances, or cascaded structures. Its response therefore depends on the spatial frequency of the incident field rather than only on the field at each individual location.} Its transfer function acts in $\mathbf{k}$-space, enabling spatial-frequency selection, differentiation, convolution, or correlation, as illustrated in Fig.~\ref{fig:wave_domain_physics}(a).

This is the basis of metamaterial analog computing, where engineered structures perform linear mathematical operations passively and in parallel as the field propagates~\cite{silva2014performing,zangeneh2021analogue}. Because early layers of learned visual encoders often implement filter banks, a nonlocal surface can perform part of this initial processing before digitisation. \textcolor{black}{The digital domain then receives a reduced representation rather than the raw field. The surface therefore acts as a hardware-constrained linear preprocessing stage rather than a general-purpose computer or a replacement for the subsequent baseband processor. The projection avoids the corresponding digital multiply--accumulate operations, although readout, conversion, calibration, surface control, and subsequent decoding still consume energy.}

\subsection{Holographic Apertures as Linear Projectors}

\textcolor{black}{A holographic aperture describes an implementation principle in which encoded reference patterns produce overlap measurements or prescribed spatial projections of an incident field. Unlike nonlocality, which describes a physical response arising from spatial interactions within a surface, holographic encoding describes how reference patterns are used to realise a projection. A nonlocal metasurface may implement a holographic projection, but the two terms are not interchangeable.} A hologram of a reference pattern couples an incident field to an output port in proportion to their overlap integral, thereby implementing an inner product. Multiple stored patterns consequently produce the projected feature vector shown in Fig.~\ref{fig:wave_domain_physics}(b). A surface containing $K$ patterns, multiplexed spatially, spectrally, or through polarisation, projects the incident field onto a designer-selected $K$-dimensional space~\cite{huang2020holographic}. Tunable elements, related to those used in reconfigurable intelligent surfaces~\cite{di2020smart}, can make the projection matrix $\mathbf{P}$ programmable at the control-electronics timescale. \textcolor{black}{A corresponding digital projection of $N$ spatial samples would require approximately $N \times K$ MACs per frame. The aperture performs this projection without executing the corresponding digital MAC operations, although readout, conversion, calibration, control, and subsequent decoding still consume energy.}

\subsection{Contribution of the Near Field} Far-field aperture responses primarily resolve direction and are described by plane-wave-like bases. \textcolor{black}{For an aperture with maximum dimension $D$ operating at wavelength $\lambda$, the conventional Rayleigh distance is approximately $2D^2/\lambda$, showing that electrically large apertures can place users and observed scenes within the radiative near field even at considerable physical distances.} In the radiative near field, wavefront curvature introduces range dependence, allowing focusing kernels indexed by three-dimensional position and giving the projection range selectivity~\cite{liu2023near}. Figure~\ref{fig:wave_domain_physics}(c) illustrates this extension from angular to position-dependent processing. For a HAPS configuration in which the aperture and operating frequency place the observed region within the Rayleigh distance, the encoder can extract position-resolved rather than only direction-resolved features. The same focusing can improve link efficiency by concentrating transmitted feature energy at the intended receiver. \subsection{Limitations of Wave-Domain Processing} A passive surface computes only linear functionals; the nonlinear operations responsible for neural-network expressiveness must remain digital. The proposed architecture therefore combines an analog projection front end with a lightweight digital decoder. Physical constraints further restrict the projection. Passivity constrains its achievable singular values, losses reduce dynamic range. \textcolor{black}{For reciprocal passive structures, independently controlling simultaneous transmit and receive transformations at the same frequency and aperture is generally difficult unless duplexing, separate modes, or nonreciprocal components are introduced.} Fabrication tolerances, control quantisation, and dispersion perturb the intended response, while reconfiguration is typically slow relative to symbol rates. The analog projection must consequently remain fixed or slowly varying within a task session. These effects must be included in the hardware--algorithm co-design of Section~IV and the conformance framework of Section~VI.

\section{The Learning Layer: Co-Designing Digital Decoders with Physical Encoders}
\label{sec:learning_layer}

Treating the aperture as the first encoder layer creates a hybrid architecture comprising a physically constrained, slowly reconfigurable linear projection, a noisy channel, and a compact digital decoder. This raises three design questions.

\subsection{Design of the Physical Projection}

The first concerns projection selection. Section~VII shows that a generic discrete cosine transform (DCT) basis performs consistently below a data-adaptive principal component analysis (PCA) projection of equal dimension. Projection choice therefore matters, although dimensionality reduction and energy concentration provide most of the benefit under the adopted link model.

The projection can be designed offline from representative data or extracted from the first linear layer of a pretrained encoder and mapped onto the aperture. The decoder is then trained for the required task and channel conditions. Alternatively, the projection can form a trainable layer in an end-to-end pipeline subject to the constraints of Section~III-D. Passivity can be imposed through row-norm constraints on $\mathbf{P}$, while hardware realisability can be enforced by mapping the optimised response onto the available meta-atom library. Quantisation-aware training can represent finite control resolution. Surface physics therefore defines the feasible projection set.

\subsection{Adaptation under a Reconfiguration Budget}

The second question concerns adaptation. The analog stage changes at the timescale of the control electronics and platform geometry, whereas the decoder can adapt more rapidly. A two-timescale architecture is therefore appropriate: the projection slowly tracks scene statistics and geometry, while the decoder compensates for channel variation, Doppler, and hardware mismatch.
This separation is compatible with the artificial intelligence/machine learning (AI/ML) framework studied by 3GPP for the NR air interface~\cite{lin2023overview}.
 A projection update can be treated as a model-update event with a defined applicability window, while decoder adaptation remains an inference-side operation.

\subsection{Distributed Semantic Processing}

The third question concerns distributed operation. Satellite or HAPS platforms may observe correlated scenes, but exchanging raw observations would burden their feeder links. Wave-domain compression instead enables the exchange of decoder gradients, model updates, or distilled feature statistics, reducing traffic and providing some data privacy. Neighbouring platforms may also use complementary physical projection bases so that the constellation collectively spans the task-relevant semantic subspace.

\section{One Aperture, Three Functions: ISAC Integration}
\label{sec:isac_integration}

\textcolor{black}{Consider a HAPS serving ground users while monitoring its coverage area. Communication patterns focus feature symbols towards terminals, sensing patterns project backscatter onto range-angle kernels, and semantic patterns project the scene onto the task-aware basis of Section~\ref{sec:learning_layer}. Time, polarisation, or spectral multiplexing allows these patterns to share the aperture and form a common feature stream.}

Two consequences are important for standardisation. First, sensing outputs can be task-level quantities throughout the processing chain. Rather than forwarding a raw point cloud or complete sensor record, the platform transmits projections onto task-relevant kernels. Feeder-link traffic is therefore governed by task requirements rather than sensor resolution, directly addressing the information-exposure question considered by the 3GPP ISAC study~\cite{3GPP_TR_22_837}. 
Second, sensing can support communication control. Near-field ISAC estimates of platform trajectory, terminal locations, and surrounding geometry can help the projection scheduler maintain focused kernels for mobile users and support tracking, mobility management, and beam alignment in dynamic NTN geometries.

Sensing thus supplies geometry to the semantic-processing layer, semantic extraction reduces sensing traffic, and near-field operation provides position-dependent resolution. The ``three musketeers'' are realised through one aperture supporting coordinated communication, sensing, and semantic-extraction patterns.

\section{Standardisation Pathways and Gaps}

Wave-domain semantic ISAC spans standards activities that currently describe the aperture as a communication or propagation component rather than a computational operator. Table~\ref{tab:standardization} summarises the current landscape and remaining gaps.

Within 3GPP, NTN work provides the transport and mobility foundation~\cite{rf1}, the AI/ML air-interface study introduces model-management concepts~\cite{lin2023overview}, and the ISAC study defines sensing use cases and service requirements~\cite{3GPP_TR_22_837}. However, no mechanism declares that a radio unit implements a particular analog projection. Interoperability would require capability signalling for supported projection families, dimensionality, bandwidth, reconfiguration latency, and implementation accuracy, allowing networks to schedule computational tasks and verify aperture behaviour. 

ITU-R identifies AI-native communication and ISAC as IMT-2030 usage scenarios~\cite{ITU2160}, but conventional evaluation still emphasises throughput, latency, and reliability. Task-oriented systems additionally require semantic key performance indicators (KPIs), such as accuracy per transmitted coefficient, task energy per inference, and semantic distortion, together with reference tasks for comparison. 
The ETSI Industry Specification Group on Reconfigurable Intelligent Surfaces (ISG RIS) has developed communication, channel, and evaluation models~\cite{ETSI_GR_RIS_003} and is addressing near-field RIS modelling and associated mechanisms~\cite{etsi007}. Nevertheless, RISs are still treated primarily as reconfigurable propagation elements. Computational apertures require descriptors for their transfer functions, projection accuracy, calibration error, and reconfiguration limits, together with TN-NTN extensions covering platform motion, Doppler, attitude variation, and time-varying near-field boundaries. 
The management plane creates a further gap. Treating a physical projection as a manageable AI/ML model would require its update and applicability events to be integrated into network control. Incorrect or malicious aperture states could affect both communication and sensing; therefore, integrity verification, authenticated updates, and safe fallback behaviour are also required.

\begin{table*}[t]
\caption{\textcolor{black}{Standardisation status and remaining gaps for wave-domain semantic ISAC in integrated TN-NTN. AI/ML denotes artificial intelligence/machine learning; HAPS denotes high-altitude platform station; and UAV denotes uncrewed aerial vehicle.}}
\label{tab:standardization}
\centering
\renewcommand{\arraystretch}{1.2}
\setlength{\tabcolsep}{4pt}
\footnotesize
\begin{tabularx}{\textwidth}{
    >{\raggedright\arraybackslash}p{0.15\textwidth}
    >{\raggedright\arraybackslash}p{0.17\textwidth}
    >{\raggedright\arraybackslash}X
    >{\raggedright\arraybackslash}X}
\toprule
\textbf{Capability} &
\textbf{Where It Sits Today} &
\textbf{What Exists} &
\textbf{What Is Missing} \\
\midrule

NTN transport and mobility
&
3GPP Radio Access Network (RAN), Release ~16 
&
NR--NTN protocol support for satellite, HAPS, and UAV links~\cite{rf1}
&
Mobility and scheduling hooks for near-field computational apertures on moving platforms
\\
\midrule

AI model lifecycle
&
3GPP RAN Working Group 1 (RAN1)~\cite{lin2023overview}
&
Air-interface AI/ML use cases, model-management concepts, and enabling mechanisms
&
Recognition of a physical analog layer as a manageable model, with applicability windows tied to reconfiguration latency
\\
\midrule

ISAC requirements
&
3GPP SA1~\cite{3GPP_TR_22_837} and RAN channel modeling
&
Use cases, requirements, and sensing channel-modeling activities
&
Task-level sensing exposure through interfaces that carry projected features rather than raw sensing data
\\
\midrule

Surface models and evaluation
&
ETSI ISG RIS~\cite{ETSI_GR_RIS_003,etsi007}
&
Communication, channel, and evaluation models, together with near-field RIS channel modelling
&
Computational-aperture capability descriptors, projection-accuracy tests, and moving-platform TN-NTN extensions
\\
\midrule

System KPIs
&
ITU-R IMT-2030~\cite{ITU2160}
&
ISAC and AI-native communication identified as usage scenarios
&
Standardized semantic KPIs, including task accuracy per symbol, task energy per inference, and semantic distortion
\\
\midrule

Trust and security
&
Cross-body gap
&
General NTN and RIS security work
&
Integrity verification for computational-aperture states, authenticated updates, and safe fallback behaviour
\\
\bottomrule
\end{tabularx}
\end{table*}

\section{Case Study: Task-Oriented Transmission Through an Analog Projection Front End}
\label{sec:case_study}

\textcolor{black}{The $8 \times 8$ image samples from the scikit-learn handwritten-digits dataset are used as abstract spatially distributed observations and are treated as input samples incident on the computational aperture. Their values are normalised observation values without physical units; they are not claimed to be calibrated electric- or magnetic-field measurements. In a practical system, the sensed quantity would either be naturally embedded in the received field or mapped onto an illuminating field by an analog transducer. The present study therefore evaluates the aperture projection and communication stages, while leaving the sensor-to-field interface and its electromagnetic calibration outside the scope of this illustrative case study.}

All schemes use fixed transmit gains, calibrated on the training set to provide equal average energy per image, $E_{\mathrm{img}}$. The receiver-noise variance, $N_0$, is fixed per transmitted coefficient; hence, a scheme transmitting $d$ coefficients has an average coefficient signal-to-noise ratio (SNR) of $E_{\mathrm{img}}/(dN_0)$. Fewer coefficients therefore concentrate the available energy into fewer channel uses.

\textcolor{black}{PCA and DCT are selected as simple and interpretable references. PCA represents a data-adaptive linear projection: its basis is obtained from the training samples and captures their dominant spatial variation. DCT represents a fixed, task-agnostic spatial transform whose basis is independent of the training data and class labels; the selected low-frequency components retain structured, slowly varying spatial content without adapting the projection to a particular dataset. Both reduced schemes retain 16 coefficients and are evaluated using the same transmitted-energy convention and decoder procedure, while the fully digital scheme serves as the reference without dimensionality reduction. These baselines isolate the effect of replacing a fixed spatial representation with a data-adaptive one; they are not an exhaustive comparison of projection methods.}

Scheme~A digitally removes the training-set mean and transmits all 64 centred samples to a decoder with 128 hidden units. Scheme~B uses a holographic front end to project the field onto the 16 leading principal components and transmits these coefficients to a 32-hidden-unit decoder. A fixed coherent reference branch cancels the projection of the training-set mean, which would otherwise consume approximately $69\%$ of the projected-field energy. \textcolor{black}{This cancellation does not require digital MAC operations at the projection stage; however, reference generation, readout, conversion, calibration, surface control, downstream decoding, and insertion loss remain outside the model's energy accounting.} 
Scheme C provides a non-data-adaptive spatial-frequency baseline implementable by a nonlocal filtering surface. It uses the same architecture and decoder as Scheme B but replaces PCA with a projection onto the 16 lowest-frequency two-dimensional DCT functions.

\begin{figure}[t]
\centering
\includegraphics[width=\columnwidth]{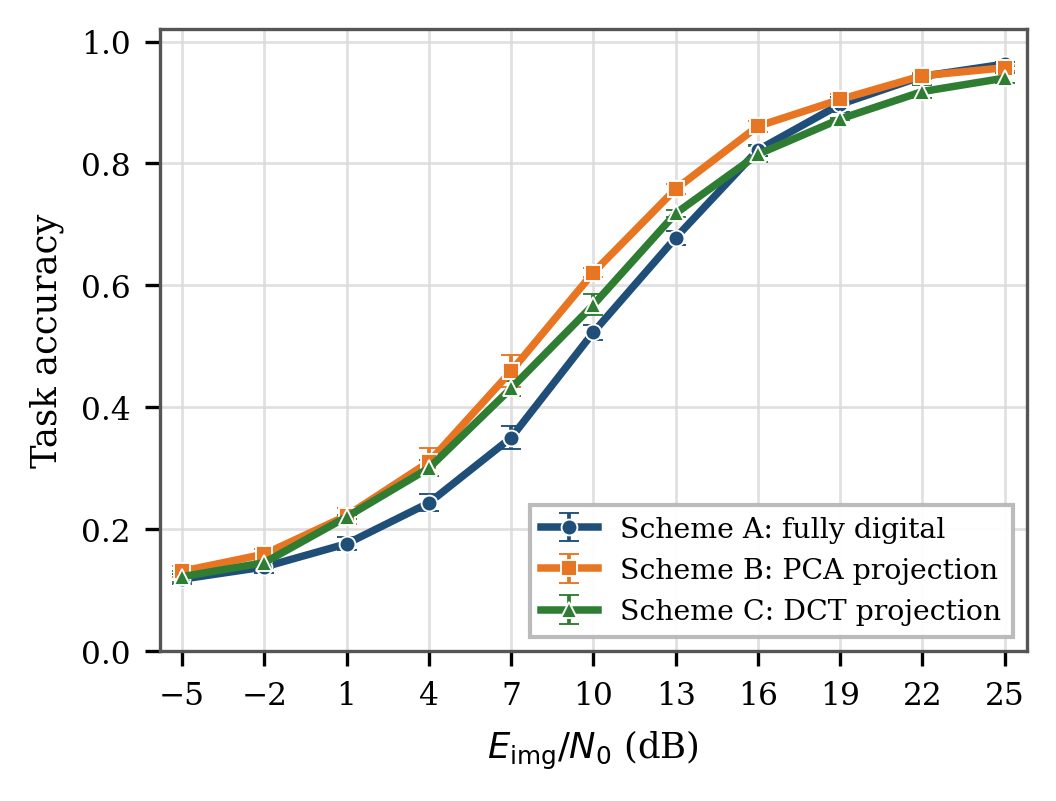}
\caption{\textcolor{black}{Numerical task accuracy versus per-image link SNR, defined as the ratio of average transmitted energy per image to the per-coefficient receiver-noise variance, so that all schemes operate at equal average transmitted energy. The fully digital, PCA, and DCT schemes use the same test procedure. Each point is the mean over five independent noise and decoder-training realisations, and the error bars denote one standard deviation across those five realisations. The input is an abstract $8 \times 8$ spatial observation and should not be interpreted as a measured electromagnetic field.}}
\label{fig:task_accuracy}
\end{figure}

\textcolor{black}{Scheme~A requires $64 \times 128 + 128 \times 10 = 9{,}472$ decoder MACs per inference. Schemes~B and~C require $16 \times 32 + 32 \times 10 = 832$ MACs. Bias additions and activation-function evaluations are excluded consistently from both counts. The wave-domain schemes therefore require approximately $9{,}472/832 = 11.4$ times fewer decoder MACs and transmit four times fewer coefficients.} Each point in Fig.~\ref{fig:task_accuracy} averages five independent noise and decoder-training realisations. 
Scheme~B outperforms Scheme~A in the simulated energy-limited range, achieving $62.1\%$ versus $52.3\%$ at $10~\mathrm{dB}$ and $86.1\%$ versus $82.2\%$ at $16~\mathrm{dB}$. The improvement results from concentrating the image energy into fewer informative coefficients. Above approximately $20~\mathrm{dB}$, the curves converge, and Scheme~A becomes marginally better at $25~\mathrm{dB}$ because its higher dimensionality becomes beneficial when noise is weak. Scheme C follows the same overall trend as Scheme B but remains approximately four to five percentage points lower across the intermediate-SNR range. This indicates that dimensionality reduction and energy concentration provide most of the gain, while adapting the physical projection to the data provides a further measurable improvement.

\textcolor{black}{The three schemes address different design questions. Scheme~A, the fully digital reference, retains the complete 64-dimensional input representation but requires the largest amount of digital processing and feature transport. Scheme~B examines whether a data-adaptive PCA representation can preserve the information needed for the classification task using only 16 coefficients. Scheme~C uses 16 fixed DCT coefficients and therefore represents a simpler projection that does not depend on the training data or task labels. The comparison is consequently intended to illustrate an accuracy-complexity tradeoff under the same average transmitted-energy convention, rather than to identify a universally optimal transform. In the experiment of Fig.~\ref{fig:task_accuracy}, the reduced representations are most useful in the energy-limited and intermediate operating ranges, where they provide higher task accuracy with fewer transmitted coefficients and lower decoder complexity; at high available energy the effect of receiver noise diminishes, and the accuracy curves converge. The preferred representation may therefore depend on the task, data distribution, propagation conditions, aperture constraints, and available digital-processing budget.}

The study assumes ideal projections, reference cancellation, equal symbol duration and bandwidth, and no binding peak-power constraint. A fabricated aperture would introduce loss, quantisation, dispersion and calibration error. Moreover, practical TN--NTN operation involves changing scene statistics and a range-, angle-, and Doppler-dependent near-field channel rather than the stationary additive white Gaussian noise (AWGN) model considered here. These limitations motivate the adaptive processing of Section~\ref{sec:learning_layer} and future measured prototypes.

\section{Open Challenges}
\label{sec:open_challenges}

Hardware realism is the first challenge. Insertion loss, dispersion, control quantisation, mutual coupling, and fabrication error perturb the implemented projection. The revised case study additionally assumes ideal coherent reference-field cancellation; practical implementations must account for reference generation, phase stability, calibration, insertion loss, and peak-power limits. Computational apertures therefore require measured prototypes and procedures for declaring and verifying their analog transfer functions.

Second, multi-task operation remains open. A slowly reconfigurable aperture cannot maintain a dedicated projection for every task. Shared bases and scheduling policies for multiple pattern families must consequently combine electromagnetic design with learning theory.

Third, robust co-design must tolerate imperfect channel and surface models, monitor semantic performance, and revert to conventional processing when a task leaves the projection applicability window. These functions should ultimately integrate with the 3GPP AI/ML model-management framework~\cite{lin2023overview}.

Fourth, evaluation remains fragmented. Agreed semantic KPIs, energy-accounting assumptions, reference channels, and task suites are needed to compare semantic ISAC proposals fairly across TN-NTN scenarios.

Finally, a computational aperture creates an attack surface for communication and sensing. Integrity verification, authenticated updates, anomaly detection, and safe recovery therefore require dedicated treatment.

\section{Conclusion}
\label{sec:conclusion}

Semantic communication, near-field operation, and ISAC can be organised around a common physical asset: a large electromagnetic aperture. Its wave-domain response performs linear semantic projections, near-field operation provides position-dependent features and focused delivery, and shared use supports communication and sensing within a common task-level architecture.

In the illustrative study, the data-adaptive wave-domain pipeline outperforms a fully digital baseline in the simulated energy-limited regime while transmitting four times fewer coefficients and using 11.4 times fewer decoder MACs; the schemes converge as the link becomes energy-rich. Moving towards deployment requires measured prototypes, computational-aperture capability declarations, semantic KPIs, moving-platform near-field channel models, and conformance procedures for spatial-frequency operators. Together, these form a concrete research and standardisation agenda for integrated TN--NTN systems.

\section{Acknowledgment}
This work was supported by the Horizon Europe Marie Skłodowska-Curie Actions (MSCA) project TRACE-V2X under Grant Agreement No. 101131204.


\ifCLASSOPTIONcaptionsoff
  \newpage
\fi

\bibliographystyle{IEEEtran}
\bibliography{IEEEabrv,Bibliography}

\vspace{-3\baselineskip}
\raggedbottom

\begin{IEEEbiographynophoto}{Shadi Danesh}
(Member, IEEE) received the M.Sc. and Ph.D. degrees in electrical-electronics and telecommunications engineering from Universiti Teknologi Malaysia in 2011 and 2015, respectively. She held Research Fellow positions at Universiti Teknologi Malaysia and Babol Noshirvani University of Technology. In 2021, she joined the Institute for Communication Systems, University of Surrey, U.K., as a Research Fellow. Her research interests include surface electromagnetics, engineered metasurfaces, dielectric resonator and reconfigurable antennas, metamaterial antennas, and satellite arrays.
\end{IEEEbiographynophoto}

\vspace{-3\baselineskip}

\begin{IEEEbiographynophoto}{Pooria Salami} received the B.Sc. degree from K. N. Toosi University of Technology in 2012, the M.Sc. degree from Amirkabir University of Technology in 2014, and the Ph.D. degree from the University of Tehran in 2020, all in electrical engineering. He is currently a Research Fellow with the Institute for Communication Systems, University of Surrey, U.K. His research interests include applied electromagnetics, metasurfaces, analog computing, computational imaging, antennas, and wavefront engineering.
\end{IEEEbiographynophoto}

\vspace{-3\baselineskip}

\begin{IEEEbiographynophoto}{Hesaam Zaravashan} is currently pursuing the Ph.D. degree with the Institute for Communication Systems (ICS), University of Surrey, Guildford, U.K., under the supervision of Dr. Mohsen Khalily. He is a member of the Surface Electromagnetics Laboratory (SEMLAB).
His research interests include photonic topological insulators, metasurfaces, and surface electromagnetics.
\end{IEEEbiographynophoto}
\vspace{-3\baselineskip}
\begin{IEEEbiographynophoto}{Abdulkadir Köse} (Member, IEEE) received the Ph.D. degree from the University of Surrey, U.K., in 2021. He is currently an Assistant Professor at Abdullah Gül University, Türkiye, and Co-Founder and CEO of Ransight Technology Ltd. He is Principal Investigator of the EU MSCA-SE TRACE-V2X project and an academic consultant for the Celtic-Next 6G-SMART project. His research interests include 5G/6G networks, O-RAN, AI-driven network automation, V2X communications, edge intelligence, and the Internet of Things.
\end{IEEEbiographynophoto}
\vspace{-3\baselineskip}
\begin{IEEEbiographynophoto}{Ayhan Akbas} received the B.Sc. and M.Sc. degrees in electrical and electronic engineering from Middle East Technical University, Türkiye, in 1991 and 1995, respectively. He held technical and managerial positions at Siemens, Sun Microsystems, and NEC. He is currently an Associate Professor at Ankara Bilim University, Türkiye. His research interests include mobile networks, 6G, O-RAN, artificial intelligence, the Internet of Things, and V2X communications.
\end{IEEEbiographynophoto}
\vspace{-3\baselineskip}

\begin{IEEEbiographynophoto}{Chuan Heng Foh} (Senior Member, IEEE) received the M.Sc. degree from Monash University in 1999 and the Ph.D. degree from the University of Melbourne in 2002. He is currently an Associate Professor at the University of Surrey, U.K. He has authored or coauthored more than 200 refereed publications. His research interests include protocol design, AI applications, AI-RAN, and network performance analysis. He is also a Senior Editor for \textit{IEEE Access}.
\end{IEEEbiographynophoto}

\vspace{-40.5\baselineskip}

\begin{IEEEbiographynophoto}{Mohsen Khalily}
(Senior Member, IEEE) is an Associate Professor in Antennas and Propagation and Head of the Surface Electromagnetics Laboratory at the University of Surrey, U.K. He has published nearly 250 articles and led research grants exceeding £6 million. He is the leading Rapporteur for the ETSI ISG RIS work item on RIS implementation and practical considerations. His research interests include surface electromagnetics, metamaterials, topological insulators, millimetre-wave and terahertz antennas, and propagation modelling. He is Subject Editor-in-Chief for Antennas and Propagation at \textit{IET Electronics Letters} and an Associate Editor for \textit{IEEE Transactions on Antennas and Propagation}, \textit{IEEE Antennas and Wireless Propagation Letters}, and \textit{Scientific Reports}.
\end{IEEEbiographynophoto}

\end{document}